# Testing randomness of series generated in Bell's experiment.

Myriam Nonaka, Mónica Agüero, Marcelo Kovalsky and Alejandro Hnilo

*CEILAP, Centro de Investigaciones en Láseres y Aplicaciones, UNIDEF (MINDEF-CONICET); CITEDEF, J.B. de La Salle 4397, (1603) Villa Martelli, Argentina.*
*email: ahnilo@citedef.gob.ar*

August 31st, 2022.

The generation of series of random numbers is an important and difficult problem. Even the very definition of "random" is difficult. Appropriate measurements on entangled states have been proposed as the definitive solution to produce series of certified randomness. However, several reports indicate that quantum-based devices show a disappointing rate of series rejected by standard tests of randomness. This problem is usually solved by using algorithms named extractors but, if the extractor were known by an eavesdropper (a situation that cannot be ruled out) the key's security in QKD setups may be menaced. We use a "toy" fiber-optic-based setup, similar to a QKD one to be used in the field, to generate binary series, and evaluate their level of randomness according to Ville's principle. Series are tested with a battery of standard statistical indicators, Hurst exponent, Kolmogorov complexity, minimum entropy, Takens' dimension of embedding, and Augmented Dickey-Fuller and Kwiatkowski–Phillips–Schmidt–Shin to check stationarity. A theoretically predicted relationship between complexity and minimum entropy is observed. The good performance of a simple method to get useful series from rejected series, reported by Solis *et al.*, is confirmed and supported with additional arguments. Regarding QKD, the level of randomness of series obtained by applying Toeplitz's extractor to rejected series is found to be indistinguishable from the level of non-rejected raw ones.

**1. Introduction.**

Series of random numbers are needed in many applications, from statistics to cryptography. Yet, randomness of a series is difficult to establish. There is not even a unanimously accepted definition of "random". There are at least two definitions of randomness that are relevant from a practical point of view [1]:

*i)* A binary series is "statistically random", uniform or Borel normal if the number of strings of 1 and 0 of different length $n$ (say, 110101 for $n=6$), matches the number one would get in a series produced by tossing an ideal coin. Other tests of statistical randomness measure the decay of the self-correlation or the mutual information, or calculate entropies. They all involve probabilities and require, in principle, the series to be stationary. The battery of tests provided to the public by the National Institute of Standards and Technology (NIST) is mostly based on this approach.

*ii)* A binary series is "algorithmically random" if there is no classical program code able to generate the series using a number of bits shorter than the series itself. Note that this definition does not involve probabilities, and applies even to non-stationary and finite series. It is directly related with the idea of *complexity* developed by Kolmogorov [2]. In few words, the complexity of a series of length $N$ is the length $K$ of the shortest program able to generate the series. If $K \approx N$ the series is incompressible, which is often considered the strongest form of randomness. The problem is that $K$ cannot be actually *computed*, for one can never be sure that there is no shorter program able to generate the series; it can be only *estimated* from the asymptotic compressibility of the series using, f.ex., the algorithm devised by Lempel and Ziv [3]. We call $Kc$ the estimated and normalized value of $K$. An algorithmically complex series is non-computable and Borel normal, but the inverses are not true [1].

Several arguments [4-7] support the idea of Quantum Certified Randomness (QCR). The series of outcomes obtained from observations of quantum systems, in particular, of entangled states in a Bell's setup, would be *intrinsically* random. A Bell's setup would provide then not only random series to be used in practice, but also a definition: *random series is what is produced by this setup*. Taking into account the mentioned difficulties, QCR idea is most appealing.

However, experimental studies show that a surprisingly high rate of those series is rejected by standard tests of statistical and algorithmic randomness [8-12]. This is caused, in principle, by instrumental imperfections. The rejection rate can be dramatically reduced by applying extractors or Pseudo Random Number Generators (Pseudo RNG), using the quantum generated series as seeds. This is acceptable to generate random series in a single place, but using extractors may be inconvenient in the case of Quantum Key Distribution (QKD), for two reasons: from the theoretical point of view, no classical algorithm can increase the "amount" of randomness in a series [7]. From the practical point of view, it is impossible to discard that the extractor is known by an eavesdropper. Summing up these two reasons, a poor level of randomness of the seed may open some vulnerability for the key's security. Besides, algorithmic randomness of quantum-produced finite series is controversial. An experimental approach was proposed to explore this problem [7,9], and was performed using estimators of $Kc$ on an almost loophole-free setup [10,11]. The observed rejection rate was as high as 25%. Observations on mixed states of different levels of entanglement found that the *non* entangled case produced series with the lowest rejection rate [13]. This case is hence optimal as a RNG but is useless for QKD, where two identical series of random numbers must be generated in two remote places.

In this paper, we study the series produced by a "toy" fiber-optic-based QKD setup, similar to the quantum channel that can be operating in the field. The aim is to decide, experimentally, whether series produced in such setup are suitable as series of random numbers (in each station) and, after the application of a standard extractor, as the key in QKD.

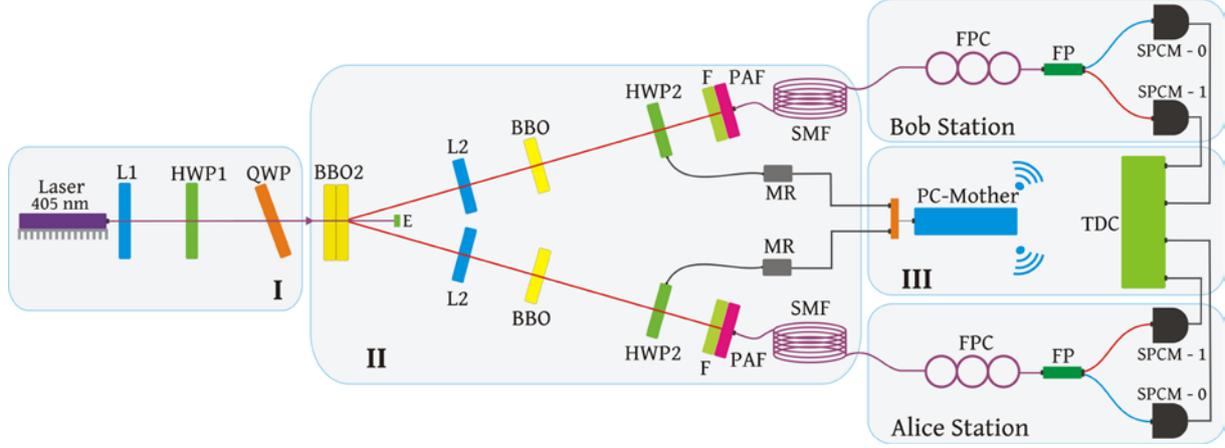

Fig. 1: Sketch of the experimental setup. Sector I: pump. Sector II: source of entangled photons. Sector III: acquisition and control. L1: plano-convex lens (f = 300 mm), HWP1: half-waveplate ($\lambda$ = 404.6 nm), QWP: quarter-waveplate ($\lambda$ = 405 nm), BBO2: two crossed BBO-I crystals ($\theta$ = 28.9º, $\varphi$ = 0º, size: 5×5×0.9 mm$^3$), L2: convex lens (f = 300 mm), BBO: ($\theta$ = 28.9º, $\varphi$ = 0º, size: 5×5× 0.45 mm$^3$), HWP2: half-waveplate ($\lambda$ =810 nm) mounted in a motorized rotation stage, MR: Servo motor controller, F: bandpass filter (($\lambda$ = 810 nm, FWHM = 10 nm), PAF: fiberport (f = 7.5 mm), SMF: 47 m of single-mode fiber coil, TDC: time-to-digital converter (4 channels), FPC: birefringence compensator, FP: fiber polarization beam splitter, SPCM: photon counting module. The PC-Mother manages the naming, opening and closing of files in the TDC through a wifi link.

In the next Section 2, the experimental setup is described, and the approach to estimate the level of randomness is explained. Observed results are discussed in Section 3.

**2. Generation of series and analysis tools.**
*2.1 Experimental setup.*

Our setup is sketched in Figure 1. It is aimed to mimic the performance of a QKD setup operating in the field. We stress that it is not loophole-free, so that our observations cannot be considered conclusive from the point of view of theoretical QCR [7].

Biphotons at 810 nm entangled in polarization in the fully symmetrical Bell state $|\phi^+\rangle$ are produced in the standard configuration using two crossed (1mm length each) BBO-I crystals and walk-off compensating crystals, pumped by a CW 40 mW diode laser at 405 nm with a measured coherence length of 20mm. The entangled photons are inserted into respective pairs of single-mode optical fibers 47 m long each. Polarization is measured with two-exit fiber-optic analyzers. Silicon avalanche photodiodes (nominal efficiency 70%) detect single photons; detections' time values are stored in a time-to-digital converter (TDC). The TDC has a nominal resolution of 10 ps, but time accuracy is reduced to ≈1 ns because of detectors' time jitter. A "mother" PC controls the servo motors that adjust the settings and manages names, opening and closing of data files in the TDC.

Fiber-optic analyzers are robust, stable and easy to align, and are natural choices for a setup operating in the field. However, their contrast (1:100) is one order of magnitude poorer than the typical one of free space ones. For this reason, the highest value for the entanglement parameter $S_{CHSH}$ that can be achieved with our setup is 2.77 (the ideal value is 2.83).

Files of time values of photon detection in the four output gates of the polarization analyzers are stored for further analysis. From a single time stamped file of raw data, two classes of series are generated, depending of the physical magnitude that is considered:

OUT (outcomes) series are made of the binary values corresponding to detection in the "0" or "1" gate in each station.

TD (time differences) series are made of the time elapsed between detections, regardless the output gate they occurred. A threshold value is used in order to transform them into binary series: we define "1" ("0") if the time difference is above (below) the threshold. The reason why we consider TD series is explained in Section 3.

Besides, from each of the TD and OUT series three different series can be extracted:

CO (coincidence) series are made of coincidences between the stations.

SO (singles only) series are made of detections that are *not* coincidences between the two stations.

AL (all detections) series are made of all detections, regardless they are coincidences or not.

Therefore, a single file of time stamped raw data produces up to six different derived series. The series made of coincident outcomes (CO+OUT) are, according to QCR, the closest to be intrinsically random. Note that, excepting for the time differences of coincidences (CO+TD), the series are different in each station.

*2.2 Some words on QKD.*

The series of outcomes produced in a setup like Fig.1, assumed random, allow the encryption of messages in a secure way using one-time-pad Vernam's cipher. This is known as Quantum Cryptography, or QKD. In few words: observers at A and B change their bases of observation of the shared entangled state in an independent and arbitrary way. Then they publicly announce the time values when each of them detected a particle, and find the coincidences. Then they announce the settings they used when coincidences were observed, *but not the outcomes* they observed. When the bases they used are parallel or orthogonal, both know the outcome observed at the other station. This list of outcomes forms a *key* that is used to encrypt the message. The outcomes obtained at the intermediate bases can be used to calculate the level of entanglement achieved. Fundamental properties of quantum physics ensure that entanglement cannot be faked by an eavesdropper; so that this method guarantees the key is known to A and B only.

That is the theory. In practice, many technical problems must be solved (and are being solved, after much effort) to make this scheme to work. Nevertheless, there is still the problem of the key's features. If the key is partially predictable, the message is vulnerable even in an otherwise technically perfect QKD setup. In fact, this was observed in one of the runs of the famous foundational Bell's experiment performed in Innsbruck [14,15] despite it was, if considered as a QKD setup, almost ideal. That result is one of the motivations of this study.

*2.3 Tests of randomness used.*

As said, randomness is not properly defined, much less can it be properly *measured*. Martin-Löf's theorem ensures that there exists a universal algorithmic test that determines if a given series is random in the typical and algorithmic senses [7]. But, the expression of this universal test is unknown. An approach at hand is as follows: a given series can be demonstrated *not-random*. This occurs when it is rejected by one of the many existing tests of randomness, both statistical and algorithmic. As the number of applied tests is increased, the result that would be obtained by applying the universal test is approached, say, asymptotically. This is known as *Ville's principle*, and is often used to evaluate the reliability of RNG in practice. A high rejection rate in the set of generated series is interpreted as a low level of randomness of the RNG. The rejection rate does not properly measure randomness, for the set of applied tests is arbitrary. Yet, common sense says that it cannot be completely unrelated with "actual randomness" (no matter how the latter is defined) if a sufficiently large and varied set of tests is used.

Following Ville's principle, and based on our previous experience, we apply the following set of tests to the series generated in the setup of Fig.1:

*NIST:* This random-checking battery includes 15 different tests. Details are publicly available in [16]. They essentially check Borel normality, hidden periodicities, and decay of mutual information. Not all tests can be applied to all series, because of series' length. A series is considered not-random if it is rejected by at least one of the *applicable* tests in the battery. In the cases analyzed in this paper, a series is always rejected by more than one test (or by none at all). We find that the tests that reject most series are the #1 (balance of 0 and 1), #4 (whether the longest series of "1" is consistent with coin-flipping randomness), #7 (number of occurrences of pre-specified target strings, aimed to detect too many aperiodic patterns), #12 (frequency of all possible overlapping *n*-bits patterns, a sort of generalized balance) and #13 (generalized path length increase compared with random walk).

*Minimum Entropy*: is defined as
$$H_{min} = -\log_2 [max_r P(r)] \qquad (1)$$
where $P(r)$ is the probability of obtaining the outcome $r$ in the series. Intuitively, it is the highest probability of guessing the next element in the series, knowing $P(r)$. In the case of the Bell's setup, $H_{min}$ is demonstrated to be bounded from below by a function of the entanglement parameter $S_{CHSH}$ [4]:
$$H_{min} \geq 1 - \log_2 [1+(2-S_{CHSH}^2/4)^{1/2}] \qquad (2)$$
for $S_{CHSH} = 2\sqrt{2}$ (maximum entanglement) $H_{min} = 1$. $H_{min}$ is bounded from above by Shannon's entropy. It is sometimes stated that $H_{min}$ measures the number of "random bits per bit", but this statement is misleading. If a series is random, it has $H_{min} = 1$ and also (because of its very definition) "one random bit per bit", but the inverse is not true. Champernowne's number or $\pi$ have $H_{min} = 1$ but are generated by an algorithm, their digits are predictable and hence, they are not random. What $H_{min} = 1$ does mean is that $P(r)$ is as uniform as it can be. In the case of binary series, $H_{min}$ is a measure of how balanced the series is. $H_{min}$ is a useful tool to help to decide whether a series is random or not, but it does *not* measure "randomness per bit" [17].

*Hurst exponent* is related to autocorrelation decay rate. It is usually named H, and normalized between 0 and 1; H>½ means the series has long range correlations, H<½ that it has strong fluctuations in the short term, while H≈½ means that it is uniform.

*Kolmogorov's Complexity* (*Kc*): As said, it cannot be actually computed, only estimated. Here we use the approach developed by Kaspar and Schuster [18] and implemented by Mihailovic [19]. This value is designed to be near to 0 for a periodic or regular series, and near to 1 for a random one. For relatively short and strongly fluctuating series, values $Kc > 1$ may occur. A relevant theoretical result is that Shannon's entropy bounds complexity from below for series produced by ergodic generators [20].

*Nonlinear analysis*: is an approach different from all statistical analyses. In a chaotic system, few dynamical variables are linked through nonlinear equations in such a way that the evolution can be unpredictable in practice. Nevertheless, this evolution involves few degrees of freedom. This is a fundamental difference with "true" random evolution, which can be thought of as requiring a very high (eventually, infinite) number of degrees of freedom to be described. Takens' reconstruction theorem [21] allows calculating the number of dimensions of the compact object (if it exists) in phase space within which the system evolves, and hence to discriminate chaos from randomness. This number is called *dimension of embedding*, $d_E$. A definite value of $d_E$, in the cases it can be reliably measured, indicates the series is not random. In some cases this approach allows the prediction of future elements of the series within a *horizon of predictability* [22]. As said at the end of Section 2.2, it was possible to predict up to 20 bits of the would-be QKD key in one of the runs of the Innsbruck experiment by using this approach. This result revealed the possibility of vulnerabilities of an unexplored type in QKD. When this result was found the setup had been dismantled, so that the cause is not known for sure; it is believed to be a drift between the clocks at the stations. In support of this belief, a much simpler time stamped experiment with a single clock failed to reveal a definite value of $d_E$ [23]. See also Section 3.

*Stationarity*: Excepting $d_E$ and $Kc$, the mentioned indicators are ultimately statistical, and hence require the series to be stationary. There are two main types of non-stationarity: *Trend-stationary*: the series' statistical parameters follow a continuous and slow evolution. Deviations from the average trend vanish as the number of elements in the series increases. By identifying and correcting the trend, the series can be made stationary again. *Unit-root*: a deviation affects the values of the statistical parameters in a permanent way through the series. Standard tests exist: Kwiatkowski–Phillips–Schmidt–Shin (KPSS) for the first type and Augmented Dickey-Fuller (ADF) for the second one. Because of the very nature of the involved hypotheses and methods, these tests do not provide definitive conclusions. Used together, they *indicate* the most probable nature of the series. KPSS tests the null hypothesis that the series is trend-stationary, against the alternative of unit-root. Obtaining "0" (1) indicates that stationarity cannot (can) be rejected. ADF tests unit-root. Obtaining "0" (1) indicates that unit-root cannot (can) be rejected. Non-stationarity does not demonstrate that the series is not random, but it casts a shadow on the reliability of the results provided by statistical tests.

## 3. Results.
*3.1 Raw series.*

Time stamped series are recorded for $S_{CHSH}$ values ranging from 2.44 to 2.73 (recall the theoretical maximum achievable for this setup is 2.77). A total of 864 series are recorded, each lasts 10 sec of continuous observation. Series of single detections have a length of 250-300 Ks (kilo samples). For optimized delay and a time window of 10 ns, series of coincidences have a typical length of 45-50 Ks; average efficiency (coincidences / singles) ranges from 18 to 22%. Discussing the features of each of these 864 series would be an extenuating task, we present here the main results. The whole set of series sum up several Gb of data, that we are glad to share upon request.

In one of the pioneering studies on the performance of quantum RNG, Solis *et al.* [8] found the series of outcomes to have a high rejection rate, and that series of time differences performed much better. In order to derive binary series from series of time differences, a threshold is defined. A time difference value above the threshold means a "1", a "0" otherwise. The question is how to find the best threshold value. In [8], the median of the distribution was chosen with satisfactory results. Here we investigate defining the threshold as the maximum of $Kc$ or $H_{min}$. As an illustration, in Figure 2, $Kc$ and $H_{min}$ are plotted as a function of the threshold value for a typical case. We find the maxima of these curves are always coincident, what establishes a single criterion to determine the threshold. Besides, this value is practically coincident with the median, concurring with [8].

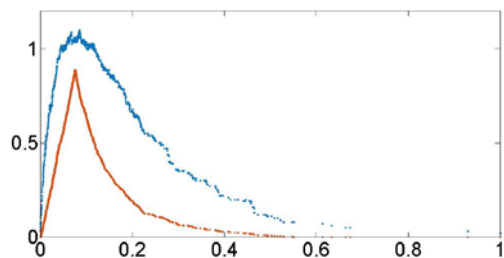

Figure 2: Example of the spectra of $H_{min}$ (orange, peaked) and $Kc$ (blue, higher) as function of (normalized) threshold value to binarize TD series. The two maxima coincide and define the threshold value. This example corresponds to a CO+TD series that passes NIST tests, $S_{CHSH}$= 2.72.

In all cases, TD series obtained in this way have higher $Kc$ and $H_{min}$ values and lower rejection rates that OUT (outcomes) ones. Therefore, TD series are more reliably random than OUT ones. The cause is that OUT series are often unbalanced because of different detectors' efficiencies or imperfect alignment. The threshold value as defined above, instead, guarantees the TD series are balanced.

No series in the whole set analyzed in this paper is found to be unit-root (i.e., we find ADF=1 in all of them), but some have KPSS=1, meaning that stationarity can be rejected. The number of suspected non-stationary series is 50% higher for OUT series than for TD ones.

All series in the whole set have H ≈ ½, meaning they show uniform decay of autocorrelation along their length. Therefore, none of them can be discarded for this reason.

The CO+OUT series have a high rejection rate (much higher than all TD series), but still lower than the other OUT series, which rejection rate is maximal

(i.e., *all* series are rejected). The CO+OUT series that are not rejected are also the ones with the highest values of H$_{min}$ (between 0.97 and 0.99) and $Kc$ (between 1.01 and 1.02) in their sets. Therefore, the series corresponding to the entangled state do have an estimated randomness higher than other related series generated at the same time, in agreement with QCR.

The CO+TD series (coincidences of time differences) have zero rejection rate, although 8 of 144 series are found not-stationary and the statistical tests are, in principle, not applicable.

The following Table summarizes the main results:

| File type | ⟨$Kc$⟩ | ⟨H$_{min}$⟩ | NIST rej. rate | KPSS=1 |
|---|---|---|---|---|
| CO+TD | 1.02 | 0.98 | 0 | 8/144 |
| CO+OUT | 0.99 | 0.80 | 0.91 | 5/144 |
| SO+TD | 1.01 | 0.99 | 0.12 | 5/144 |
| SO+OUT | 1.00 | 0.86 | 1 | 15/144 |
| AL+TD | 1.01 | 0.99 | 0.24 | 5/144 |
| AL+OUT | 1.00 | 0.88 | 1 | 7/144 |

Table: Summary of results, averaged over the whole set of raw series analyzed in this paper (864 series) unless indicated otherwise. F.ex: 8/144 in the CO+TD line and KPSS=1 column means that 8 out of 144 series have KPSS=1. See the main text for details.

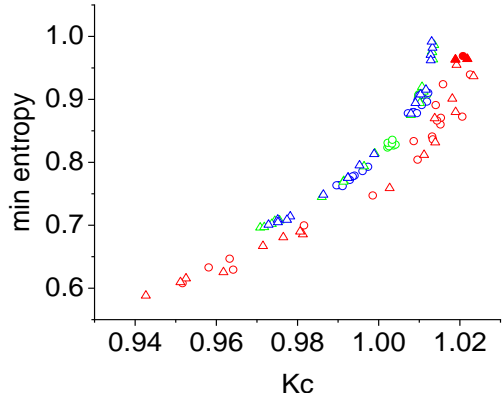

Figure 3: An example of the observed relationship between Complexity ($Kc$) and mínimum entropy (H$_{min}$) for OUT series. Red: CO, Blue: SO, Green: AL. Circles: series recorded in station A, triangles: station B. Full (open) signs mean the series is not rejected (rejected) by the NIST battery. In this run S$_{CHSH}$ = 2.66.

We measure H$_{min}$ for several values of S$_{CHSH}$ in order to observe the relationship in eq.2, but the result is disappointing. Measured H$_{min}$ is always much larger than the bound in eq.2, its variation with S$_{CHSH}$ is small and showing no definite trend. In other words: experimentally obtained series (both TD and OUT) are always much more uniform than it is guaranteed by eq.2. F.ex., for the highest observed value S$_{CHSH}$ = 2.73, the bound is H$_{min}$ ≥ 0.546 but the average measured value of H$_{min}$ is 0.90, and none of the recorded series has a H$_{min}$ value smaller than 0.74.

An important theoretical result is that classical entropy (and hence, H$_{min}$) puts a minimum bound to $Kc$ if the generator of the series is ergodic [20]. This relationship is clearly seen in the data of OUT series, see Figure 3. In fact, estimated $Kc$ is always higher than the bound put by H$_{min}$. This holds true even for the few series that are not-stationary, which are the natural candidates to be produced by a non-ergodic generator.

For the TD series recorded in the same run, the bound still holds, but the relationship between $Kc$ and H$_{min}$ is not visible, see Figure 4. Note the higher values of both magnitudes, compared with Fig.3.

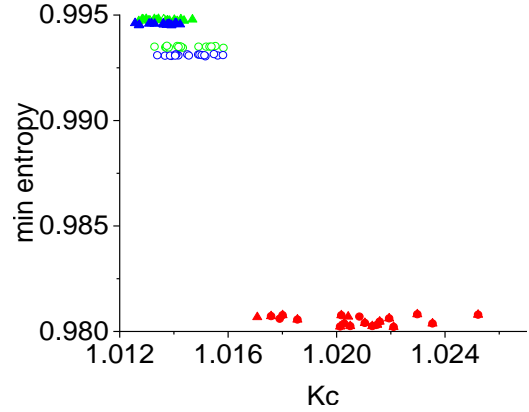

Figure 4: the same as Fig.3 but for TD series, same run and color and signs' codes. Note the different ranges in both axes.

Takens' method finds 5 series with a definite value of d$_E$ = 9. All of them are CO+TD series recorded during the same experimental run with S$_{CHSH}$ = 2.72, which is one of the highest recorded values of entanglement in this study. Noteworthy, all these series have coincident values of the Lyapunov exponents. The highest positive Lyapunov is 0.13, what roughly corresponds to a horizon of predictability of 7 to 8 elements in the series. Nevertheless, these series are random according to NIST (recall the rejection rate for the set of CO+TD series is zero). Notably, one of the CO+OUT corresponding series is random according to NIST too. However, the existence of an attractor implies they are not truly random. None of these series appear in any special position in the plots analogous to Figure 3 and 4 (which display data of a different run).

We attempt to reconstruct the attractor with d$_E$=9 that produced both TD and OUT series not rejected by NIST. That may allow us to predict the series up to the horizon of predictability, as it was done in [15] for one of the runs in the Innsbruck experiment. This approach requires reconstructing the attractor for each sub-series corresponding to each possible outcome (i.e., 00, 01, 10, 11). Recall that in QKD the time values of coincidences are public, what are kept secret are the outcomes. Once the attractor for each sub-series is reconstructed, the predicted time value (among the four possible ones) that is closest to the publicly announced time value defines the guessed outcome. In the file in Innsbruck, the attractor was successfully reconstructed for all the sub-series, despite some of them were as short as 2 Ks. Here, instead, none of the sub-series has a reliable value of d$_E$ (the result d$_E$=9 is obtained for the *complete* series only), so that no prediction of the key is possible. This result is consistent with the mentioned belief, that the attractor observed in [15] was caused by a drift between the clocks.

Anyway, observing a definite value of $d_E$ is quite extraordinary. In the set of data of the Innsbruck experiment, it was observed in only one file among 365. Here, in 5 among 432, and all of them in the same run, that is, near to each other in real time. The cause of the attractor in our case cannot be a drift between the clocks for there is only one clock, but it may be some unusual event (say, a small earthquake).

*3.2 Extracted series.*

The useful series for QKD are the CO+OUT ones, but only 9% of them pass NIST tests (see Table 1). This is due to experimental imperfections, mostly differences of detectors' efficiencies and imperfect alignment. These features are common to many QKD setups, and are to be expected for a device operating in the field. The usual solution is to apply a standard extractor of randomness.

We choose the well-known Toeplitz hashing randomness extractor [24], implemented with the "toeplitz" function of Matlab (R2016b), which can run in a standard PC. The method is based on matrix multiplication. The first row and the first column of a $m \times n$ matrix are made of elements of the series to be extracted. The remaining elements of the matrix are determined from these data according to a constructing one-way algorithm. Hence, the total number of raw bits needed to build the matrix is $n+m-1$. Afterwards, the matrix is multiplied by a $n \times 1$ seed matrix made of other $n$ data of the raw series. The resulting $m \times 1$ matrix is the extracted series, which is hence shorter than the raw one. The first step to apply this extractor is to choose the two integers $m$ and $n$, taking into account the length of the available raw series and the capacity of calculus. In our case we choose $m = n = 2^{14} = 16384$, which is therefore the length of all the extracted series.

Now we take the position of an eavesdropper who knows the extractor used and tries to distinguish "raw" from "extracted" series in order to find some vulnerability. None of the extracted series is rejected by statistical tests; extractors are designed to achieve precisely that. In general, we find no significant difference between raw and extracted series. The few differences we find are commented next

Extracted series have nearly the same average H value than raw random ones, yet, the extracted series tend to reach values farther from ½ than the raw ones. This indicates the presence of short and long range fluctuations in the extracted series more often than in the raw ones. Dispersion of H values is 0.024 in the extracted series, against 0.015 in the raw ones. In concordance with this result, the average value of $Kc$ in the extracted series increases from 0.99 (see Table 1) to slightly above 1 ($\langle Kc \rangle = 1.03$), what is characteristic of strong fluctuating series.

Extracted series also have larger values of $H_{min}$, in fact $\langle H_{min} \rangle$ increases from 0.80 (see Table 1) to 0.96. This is a typical consequence of applying an extractor.

These differences do not allow separating "raw" from "extracted" series, much less to predict the series and break the key. We conclude that the series obtained by using Toeplitz's extractor, in our original set of rejected CO+OUT series, are suitable for QKD.

**Conclusions.**

We analyze the level of randomness, according to Ville's principle, of series of outcomes and time differences obtained in a time-stamped all-fiber-optical Bell's experiment, aimed to mimic a QKD setup used in the field. As already stated, our setup is not loophole-free. Therefore, from the point of view of QCR theory, our results cannot be considered conclusive. They must be considered as preliminary or partial results. Nevertheless, it is to be noted that reaching the loophole-free condition in a setup operating in the field is a formidable challenge, probably out of reach of current technology.

If the purpose is to build a RNG, the series of time differences of coincidences is the best choice. They all have high values of $Kc$ and $H_{min}$ and none is rejected by NIST tests. The best threshold value to binarize the series is the maxima of the $Kc$ and $H_{min}$ spectra, which are practically coincident with the median of the distribution in all analyzed cases. The procedure of using time difference series can be thought of as a sort of "physical extractor".

If the purpose is QKD instead, series of outcomes must be used. These series have acceptable values of $Kc$ and $H_{min}$ but a too high rejection rate by NIST tests. We then apply Toeplitz's algorithm to get extracted series. These series are indistinguishable from the original non-rejected or "raw random" ones so that they are, in principle, suitable for QKD. Of course, there always remains the possibility that new tests find some regularity in the extracted series making them unsuitable. F.ex, the use of neuronal networks has been recently proposed to predict series generated by quantum RNG [25].

An interesting result of our study is the observation of the theoretically predicted relationship between $Kc$ and entropy. The prediction applies to series produced by ergodic generators, but observation indicates that it can apply also to series that are non-stationary, and hence, presumably produced by a non-ergodic source.

**Acknowledgments.**

Many thanks to Prof. Dragutin Mihailovic for his help to use his algorithms to estimate Kolmogorov complexity and to interpret their outputs. This work received support from the grants N62909-18-1-2021 Office of Naval Research Global (USA), and PIP 2017 0100027C CONICET (Argentina).

**References.**